\documentclass[preprintnumbers,
a4paper,
superscriptaddress,
prd,nofootinbib,floatfix,12pt,
]{revtex4}

\usepackage{graphicx,amssymb,amsmath,color}
\usepackage{hyperref}
\usepackage{color}
\usepackage{setspace}

\def\beq{\begin{equation}}
\def\eeq{\end{equation}}
\def\ceq{\end{equation} \begin{equation}}
\def\bea{\begin{eqnarray}}
\def\eea{\end{eqnarray}}
\def\bei{\begin{itemize}}
\def\eei{\end{itemize}}
\def\bmat{\begin{matrix}}
\def\emat{\end{matrix}}
\def\ble{\begin{flushleft}}
\def\ele{\end{flushleft}}
\def\={\,=\,}
\def\+{\,+\,}
\def\-{\,-\,}

\newcommand{\Fig}[1]{Fig.~\ref{#1}}
\newcommand{\Eq}[1]{Eq.~(\ref{#1})}
\newcommand{\Sec}[1]{Sec.~\ref{#1}}

\begin{document}

\title{Stop and Sbottom LSP with R-parity Violation}

\author{Eung Jin Chun}
\email{ejchun@kias.re.kr}
\affiliation{Korea Institute for Advanced Study, Seoul 130-722, Korea}

\author{Sunghoon Jung}
\email{nejsh21@gmail.com}
\affiliation{Korea Institute for Advanced Study, Seoul 130-722, Korea}

\author{Hyun Min Lee}
\email{hmlee@cau.ac.kr}
\affiliation{Department of Physics, Chung-Ang University, Seoul 156-756, Korea}

\author{Seong Chan Park}
\email{s.park@skku.edu}
\affiliation{Department of Physics, Sungkyunkwan  University, Suwon 440-746, Korea}

\begin{abstract} \vspace{3mm} \baselineskip=16pt
Considering a third-generation squark as the lightest supersymmetric particle (LSP), we investigate R-parity violating collider signatures with bilinear LH or trilinear LQD operators that may contribute to observed neutrino masses and mixings.
Reinterpreting the LHC $7+8$\,TeV results of SUSY and leptoquark searches, we find that third-generation squark LSPs decaying 
to first- or second-generation leptons are generally excluded up to at least about 660 GeV at 95\%C.L.. 
One notable feature of many models is that sbottoms can decay to top quarks and charged leptons that lead to a broader invariant mass spectrum and weaker collider constraints. More dedicated searches with $b$-taggings or top reconstructions are thus encouraged.
Finally, we discuss that the recently observed excesses in the CMS leptoquark search can be accommodated by the decay of sbottom LSPs in the LQD$_{113+131}$ model.

\end{abstract}

\preprint{KIAS-P14051}

\maketitle

\newpage

\baselineskip=15pt
\tableofcontents

\baselineskip=18pt

\section{Introduction}

Supersymmetry  (SUSY) has been considered as a leading candidate for physics beyond the Standard Model because it provides a natural framework to stabilize the weak scale against huge quantum corrections. 
The CMS and ATLAS collaborations of the LHC experiment have been performing a broad range of searches for SUSY in various channels.  After the LHC Run-1 with the $\sqrt{s}=7, 8$ TeV collision energies, the first two generation squarks and gluinos are already excluded up to $1\sim2$  TeV and the third generation squarks up to $400\sim 700$ GeV depending on various search channels with R-parity conservation (RPC) or violation (RPV)~\cite{susy-summary}.
Among three generations of squarks, the third generation squarks are of particular interest, as they contribute significantly to the Higgs mass through loop corrections, and thus direct stop/sbottom searches at the LHC are motivated.  

As is well-known, the Standard Model gauge invariance allows bilinear (LH) and trilinear (LLE, LQD) lepton-number ($L$) violating operators as well as trilinear (UDD) baryon-number ($B$) violating operators in the renormalizable superpotential:
\bea
W_{\rm RPV} = \epsilon_i \mu L H_u  + \lambda_{ijk} L_i L_j E^c_k+ \lambda'_{ijk} L_i Q_j D^c_k 
+ \lambda''_{ijk} U^c_i D^c_j D^c_k
\eea
where $\mu$ denotes the supersymmetric mass parameter of the Higgs bilinear operator $H_u H_d$. 
Simultaneous presence of $\lambda'$ and $\lambda''$ makes proton unstable and thus has to be avoided.
The proton stability may be ensured by imposing various discrete symmetries~\cite{ibanez92}.
One of them is the standard R-parity forbidding all of the above operators.\footnote{Note that the dimension-5  $B$ and $L$ violating operator $LQQQ$, which is R-parity even, is assumed to be highly suppressed in addition.} 
Another popular options are to consider the B-parity and L-parity forbidding only $B$ and $L$ violating operators, respectively. 
The B-parity has an attractive feature that the allowed $L$ violating operators could be the origin of  tiny neutrino masses~\cite{hall84}.
 
 \medskip
 
Motivated by these, we investigate signatures of stop/sbottom LSP directly decaying into a quark and a lepton through either the bilinear LH or trilinear LQD couplings which can contribute to the observed neutrino masses and mixing.
One of the search channels for such RPV stop/sbottom is the conventional leptoquark search \cite{BRWmodel}  which have been looked for at the HERA~\cite{HERA}, and more recently at the LHC~\cite{cms-LQ,cms-LQ2,atlas-LQ,cms-LQ3}. In this paper,
we study various prompt multilepton and/or multijet signatures of the stop/sbottom LSP with the LH or LQD RPV to constrain the stop/sbottom mass  combing all the relevant current LHC results not only from the leptoquark search but also from the RPC stop/sbottom as well as RPV multilepton searches.  Our RPV models can have various types of couplings such as LH$_i$ and  LQD$_{ij3, i3k}$,
and the interpretation of data in terms of these variant models can be different.  The $L$ violating RPV signatures of stop have been studied earlier in  Refs.~\cite{bartl96}, and more recently in Ref.~\cite{evans12}. Leptoquark signatures of stop/sbottom LSP 
have also been explored recently in Ref.~\cite{marshall14} in the context of a bilinear spontaneous RPV model.

The CMS has also reported excesses in the leptoquark mass range of $600-700$ GeV in both $eejj$ and $e\nu jj$ channels with $2.4\sigma$ and $2.6\sigma$, respectively~\cite{cms-LQ}.  These excesses are characterized by jets from non-$b$ quarks. 
On the other hand, no similar excess is observed in $\mu\mu(\nu) jj$ and $\tau\tau(\nu) jj$ channels. 
It is attempting to see if such observed signatures are understood by any of RPV stop/sbottom LSP decay processes. 
Interestingly, these excesses can be accommodated in the sbottom LSP scenario with appropriate LQD operators. 
This may  have some implication on the other stop/sbottom masses from the electroweak precision data (EWPD). One can find other attempts to explain the excess in Refs.~\cite{Bai:2014xba}.

This paper is organized as follows.
We start by deriving the stop/sbottom RPV vertices arising from the LH and LQD couplings and reviewing their implication to the neutrino mass matrix, and then we set up  benchmark models specified by various LH and LQD couplings in \Sec{sec:models}.
Various LHC 7+8 TeV results are reinterpreted to constrain these benchmark models in \Sec{sec:bounds}.
Several qualitatively different models are considered and dedicated searches are proposed. 
The comparison with RPC model constraints is another useful result of this paper.   \Sec{sec:LQ} addresses the issue 
of accommodating  the recently observed mild excesses in the CMS leptoquark  searches in our context, and its possible implication to 
EWPD constraints. Finally we conclude in \Sec{sec:summary}.


\section{Models with LH and LQD RPV}
\label{sec:models}

\subsection{General Consideration}

As mentioned, we consider the LH and LQD operators relevant for the stop and sbottom LSP decays:
\beq
W_{\rm RPV} = \epsilon_i \mu L_i H_u + \lambda'_{ijk} L_i Q_j D^c_k \,.
\eeq

\underline{LH:}\, Let us first derive the stop and sbottom couplings arising from the bilinear LH RPV. 
For this, we need to include also soft SUSY breaking bilinear terms,
\beq
V_{\rm soft, LH} \= B_i  L_i H_u \+ m^2_{L_i H_d} L_i H_d^\dagger  \+ h.c.,
\eeq
which generate the vacuum expectation value (vev) of a sneutrino field $\tilde \nu_i$ parameterized as $\langle {\tilde \nu}_i \rangle \, \equiv \, a_i \langle H_d \rangle$ with $a_i = (B_i t_\beta + m^2_{L_i H_d} )/m^2_{\tilde \nu_i}$. Here $t_\beta$ is the ratio between two Higgs vevs: $t_\beta = \langle H_u^0 \rangle/ \langle H_d^0 \rangle$
The bilinear couplings $\epsilon_i$ and $a_i$ induce mixing masses between neutrinos (charged leptons) and neutralinos (charginos) and thereby non-vanishing neutrino masses as well as effective RPV couplings of the stop and sbottom LSP of our interest.  To see this, it is convenient to diagonalize away first these mixing masses 
as discussed in Ref.~\cite{chun02}. The relevant approximate diagonalizations valid in the limit of $\epsilon_i, a_i \ll 1$ are 
collected in Appendix~\ref{app:diag}. 
After these diagonalizations, we get the following RPV vertices of stops:
\begin{eqnarray}
-{\cal L} &=&  \tilde{t}_L \bar{t} \left( \kappa^t_{L \nu_i} P_L + \kappa^t_{R \nu_i} P_R\right) \nu_i 
               +  \tilde{t}_R \bar{t} \left( \rho^t_{L  \nu_i} P_L + \rho^t_{R  \nu_i} P_R\right) \nu_i + h.c. \\ 
                &+&  \tilde{t}_L \bar{b} \left( \kappa^t_{L  e_i} P_L + \kappa^t_{R  e_i} P_R\right) e_i 
               +  \tilde{t}_R \bar{b} \left( \rho^t_{L  e_i} P_L + \rho^t_{R  e_i} P_R\right) e_i  + h.c.,\\
\mbox{where} 
&& \kappa^t_{L\nu_i} =  y_t c^N_4 \xi_i c_\beta  \,,  ~~~~~~~~~~~~
\kappa^t_{R \nu_i} = ({\sqrt{2}\over6} g' c^N_1 + {1\over\sqrt{2}} g c^N_2)  \xi_i c_\beta, \\
&& \rho^t_{L\nu_i} =  {2\sqrt{2}\over3} g' c^N_1 \xi_i c_\beta  \,,  ~~~~~~~
\rho^t_{R\nu_i} = y_t c^N_4  \xi_i c_\beta,   \\
&& \kappa^t_{L e_i} =  -y_b c^L_2 \xi_i c_\beta +  y_b \epsilon_i \,,  ~~
\kappa^t_{R e_i} = g {m_i^e \over F_C} c^R_1 \xi_i c_\beta, \\
&& \rho^t_{L e_i} =  0 \,, ~~~~~~~~~~~~~~~~~~~~~~
\rho^t_{R e_i} =  -y_t {m^e_i \over F_C} c^R_2 \xi_i c_\beta. 
\end{eqnarray}
Similarly, the sbottom RPV vertices are given by
\begin{eqnarray}
-{\cal L} &=&  \tilde{b}_L \bar{b} \left( \kappa^b_{L  \nu_i} P_L + \kappa^b_{R  \nu_i} P_R\right) \nu_i 
               +  \tilde{b}_R \bar{b} \left( \rho^b_{L  \nu_i} P_L + \rho^b_{R  \nu_i} P_R\right) \nu_i + h.c. \\ 
                &+&  \tilde{b}_L \bar{t} \left( \kappa^b_{L  e_i} P_L + \kappa^b_{R  e_i} P_R\right) e_i 
               +  \tilde{b}_R \bar{t} \left( \rho^b_{L  e_i} P_L + \rho^b_{R  e_i} P_R\right) e_i  + h.c., \\
\mbox{where}
&&\kappa^b_{L\nu_i} = y_b c^N_3 \xi_i c_\beta - y_b \epsilon_i \,, ~~~ 
\kappa^b_{R\nu_i} =  ({\sqrt{2}\over6} g' c^N_1 - {1\over\sqrt{2} } g c^N_2 ) \xi_i c_\beta, \\
&&\rho^b_{L\nu_i} = -{\sqrt{2}\over 3} g' c^N_1 \xi_i c_\beta   \,, ~~~~~
\rho^b_{R\nu_i} = y_b c^N_3 \xi_i c_\beta - y_b \epsilon_i,   \\
&& \kappa^b_{L e_i} =  -y_t {m^e_i \over F_C} c^R_2 \xi_i c_\beta \,, ~~~~~ 
\kappa^b_{R e_i} = g {m_i^e \over F_C} c^R_1 \xi_i c_\beta,  \\
&& \rho^b_{L e_i} =  0 \,, ~~~~~~~~~~~~~~~~~~~~~
\rho^b_{R e_i} =  -y_b c^L_2 \xi_i c_\beta + y_b \epsilon_i. 
\end{eqnarray}

\underline{LQD}: It is straightforward to get the stop and sbottom RPV vertices coming from 
the trilinear RPV couplings, $\lambda'_{ijk}$ with $j$ or $k=3$: 
\begin{eqnarray}
-{\cal L} &=& \lambda'_{i33} \left( 
\tilde{b}_L \bar{b} P_L \nu_i + \tilde{b}_R \bar{b} P_R \nu_i 
-\tilde{t}_L \bar{b} P_L e_i - \tilde{b}_R \bar{t} P_R e_i
 \right) + h.c. \\
&+& 
\lambda'_{ij3} \left(  \tilde{b}_R \bar{d}_j P_R \nu_i - \tilde{b}_R \bar{u}_j  P_R e_i   \right)  + h.c. \\
&+&
\lambda'_{i3k} \left(  \tilde{b}_L \bar{d}_k P_L \nu_i - \tilde{t}_L \bar{d}_k  P_L e_i 
  \right) + h.c..
\end{eqnarray}

\medskip

When the LH and LQD PRV are allowed, their couplings can contribute to generate neutrino mass matrix components
respectively at tree and one-loop (see Fig.~1) as follows: 
\begin{eqnarray} 
\label{nu-tree}
m^{\rm tree}_{\nu, ij} &=& {M_Z^2 \over F_N} \xi_i \xi_j c^2_\beta \,,   \\
\label{nu-loop}
m^{\rm loop}_{\nu, ij} &=&  \sum_{k=1}^3 
{3\over 16\pi^2} (\lambda'_{ik3} \lambda'_{j3k} + \lambda'_{i3k} \lambda'_{jk3})
 { m_{d_k} m_b X_b \over m^2_{\tilde b_2} - m^2_{\tilde b_1}} \ln{   m^2_{\tilde b_2} \over m^2_{\tilde b_1}}, 
\end{eqnarray}
where  $m_b X_b$ is the sbottom mixing mass-squared and only sbottom contributions are included assuming $m_{\tilde b} \ll m_{\tilde d_k}$ for $k=1,2$. 
A complete 1-loop calculation can be found in Refs.~\cite{chun99,chun02}. 
In the case of the neutralino LSP, the RPV signatures correlated with the neutrino mixing angles have been extensively studied~\cite{neutrino1,neutrino1.1,neutrino1.2} as well as in the split SUSY~\cite{neutrino2}. 
Similar studies are worthwhile in the case of the stop/sbottom LSP as well. We leave this issue as a future work. 

From the expressions in Eqs.~(\ref{nu-tree}, \ref{nu-loop}), 
the LH and LQD couplings are constrained by the measured values of tiny neutrino masses. 
As a rough estimate, the following bilinear and trilinear couplings are required to generate 
the neutrino mass components of $m_{\nu, ii} = 0.01$ eV: 
\bea
\label{xi-mnu}
&& |\xi_i c_\beta| \approx 10^{-6}\,, \\
\label{lambda-mnu}
&&|\lambda'_{ik3}\lambda'_{i3k}|^{1/2} \approx 3.4\times10^{-3} \sqrt{m_d \over m_{d_k}}, 
\eea
taking $F_N=X_b = 
\sqrt{ (m^2_{\tilde b_2} - m^2_{\tilde b_1})/ \ln(m^2_{\tilde b_2}/m^2_{\tilde b_1})}= 1$ TeV. 
These coupling sizes are small enough that they do not affect production rates and do not make resonances broader than experimental resolutions so that collider physics is mostly independent on them. Nevertheless, they are large enough to allow prompt decays of squark LSPs.

\begin{figure}[t] 
\includegraphics[width=0.49\textwidth]{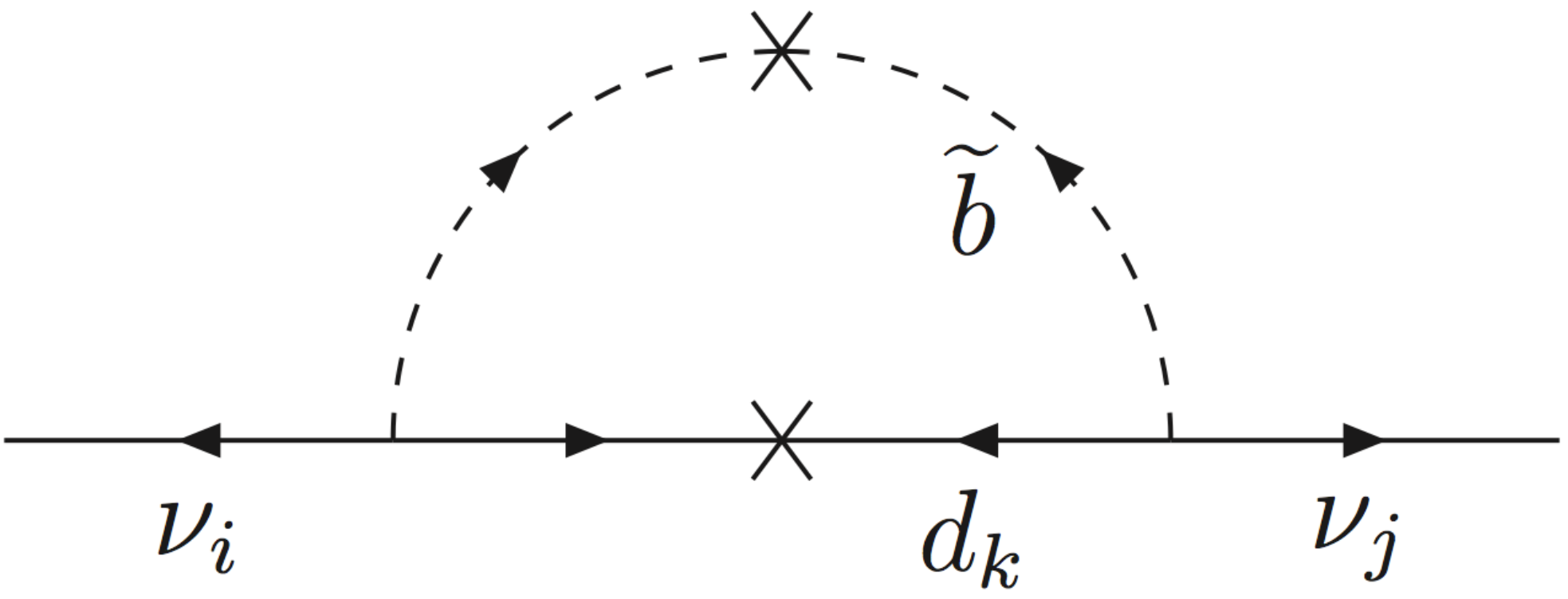}
\caption{Feynman diagrams responsible for neutrino mass generation, $m_{\nu,ij}$, through light sbottoms and LQD couplings $\lambda^\prime_{ik3} \lambda^\prime_{j3k}$.}
\label{fig:neutrinomass}
\end{figure}

\subsection{Benchmark Models}

We now introduce three benchmark models. Sbottom and stop LSPs decay to either first- or second-generation leptons. Model names imply the involved RPV interactions and subscripts imply lepton and/or quark generations.

In the presence of the mixing between left-handed and right-handed stops/sbottoms, 
we can write the stop/sbottom mass eigenstates, ${\tilde q}_1$ and ${\tilde q}_2$  with $q=t, b$:
\bea
{\tilde q}_L &=& \cos\theta_{\tilde q}\,{\tilde q}_1 - \sin\theta_{\tilde q}\,{\tilde q}_2, \\
{\tilde q}_R &=& \sin\theta_{\tilde q}\,{\tilde q}_1 + \cos\theta_{\tilde q}\,{\tilde q}_2,
\eea
where $\theta_{\tilde q}$ is the squark mixing angle. We are interested in the RPV vertices of the lightest stop ($\tilde t_1$) 
or sbottom ($\tilde b_1$).

\medskip
\underline{LH$_i$}: Stop and sbottom decay modes are  $\tilde b_1 \to e_i t, \nu_i b$ and $\tilde t_1 \to e_i b, \nu_i t$,
 and the
branching fraction for the charged lepton modes are given by (ignoring top and bottom masses)
\bea
\beta_{\tilde b} &\equiv & {\rm BR}(\widetilde{b}_1 \to e_i t) \ \approx 
{ \sin^2\theta_{\tilde b} |\rho^b_{R e_i}|^2 \over
|\kappa^b_{L \nu_i}|^2 + \cos^2\theta_{\tilde b} |\kappa^b_{R \nu_i}|^2 + \sin^2\theta_{\tilde b} |\rho^b_{L \nu_i}|^2   + \sin^2 \theta_{\tilde b} |\rho^b_{R e_i}|^2}, 
\\
\beta_{\tilde t} &\equiv& {\rm BR}(\widetilde{t}_1 \to e_i b) \ \approx
{ \cos^2\theta_{\tilde t} |\kappa^t_{L e_i}|^2 \over
|\kappa^t_{L \nu_i}|^2 + \cos^2\theta_{\tilde t} |\kappa^t_{R \nu_i}|^2 
+ \sin^2\theta_{\tilde t} |\rho^t_{L \nu_i}|^2   + \cos^2\theta_{\tilde t} |\kappa^t_{L e_i}|^2  },
\label{eq:bt-lh} \eea
where we neglect the terms suppressed by $m^e_i/F_C$.  
As the stop or the sbottom is the LSP, it is expected to have $M_Z \ll \mu$ and thus 
$|c^N_{3,4}, c^{L,R}_2 | \ll |c_{1,2}^N, c^{L,R}_1|$, which leads to 
\bea
\beta_{\tilde b} & \approx& 
{ \sin^2\theta_{\tilde b} |y_b \epsilon_i |^2 \over
\left[\cos^2\theta_{\tilde b} |{\sqrt{2} \over 6}g' c^N_1 - {1\over \sqrt{2}} g c^N_2|^2 
+ \sin^2\theta_{\tilde b}  |{\sqrt{2} \over 3} g' c^N_1 |^2 \right] |\xi_i c_\beta|^2 
+  (1 + \sin^2\theta_{\tilde b} ) |y_b \epsilon_i|^2  },
\\
\beta_{\tilde t} &\approx&
{ \cos^2\theta_{\tilde t} |y_b \epsilon_i |^2 \over
\left[\cos^2\theta_{\tilde t} |{\sqrt{2} \over 6}g' c^N_1 + {1\over \sqrt{2}} g c^N_2|^2 
+ \sin^2\theta_{\tilde t}  |{2\sqrt{2} \over 3} g' c^N_1 |^2  \right] |\xi_i c_\beta|^2 
 + \cos^2\theta_{\tilde t} |y_b \epsilon_i|^2  }\,.
\eea
Note that 
the LH model becomes effectively equivalent to the LQD$_{i33}$ model with $\lambda'_{i33} \equiv \epsilon_i y_b$ 
(see below) in the limit of vanishing $\xi_i$.

\medskip

\underline{LQD$_{i33}$}: Only $\lambda^\prime_{i33} \ne 0$ is assumed to allow the decay modes $\tilde b_1 \to e_i t, \nu_i b$ or $\tilde t_1 \to e_i b$. Thus, the sbottom and stop decay branching ratios for the charged lepton modes are 
\beq
\beta_{\tilde b} \, \equiv \, {\rm BR}(\widetilde{b}_1 \to e_i t) \= \frac{\sin^2\theta_{\tilde b}}{1+\sin^2\theta_{\tilde b}}, 
\ceq
\beta_{\tilde t} \, \equiv \, {\rm BR}(\widetilde{t}_1 \to e_i b ) \= 1\,. 
\eeq

\medskip
\underline{LQD$_{ij3+i3j}$}: Only $\lambda^\prime_{ij3,\,i3j} \ne 0$ is assumed to allow  
$\tilde b_1 \to e_i u_j, \nu_i d_j$ or $\tilde t_1 \to e_i d_j$. 
The sbottom and stop branching ratios for the charged lepton modes are
\bea
\beta_{\tilde b} &\equiv& {\rm BR}(\widetilde{b}_1 \to e_i u_j) \= \frac{\sin^2\theta_{\tilde b}|\lambda'_{ij3}|^2 }{\cos^2\theta_{\tilde b}|\lambda'_{i3j}|^2+2\sin^2\theta_{\tilde b} |\lambda'_{ij3}|^2}, 
\label{eq:beta-leptoquark}
\\
\beta_{\tilde t} &\equiv& {\rm BR}(\widetilde{t}_1 \to e_i d_j) \= 1\,.
\eea

The first two models, LH$_i$ and LQD$_{i33}$,  involve heavy quarks (tops and bottoms) in the final states while only light quarks are produced in the LQD$_{ij3+i3j}$ model.

\section{LHC Searches and Bounds}
\label{sec:bounds}

\begin{figure}[t] 
\includegraphics[width=0.49\textwidth]{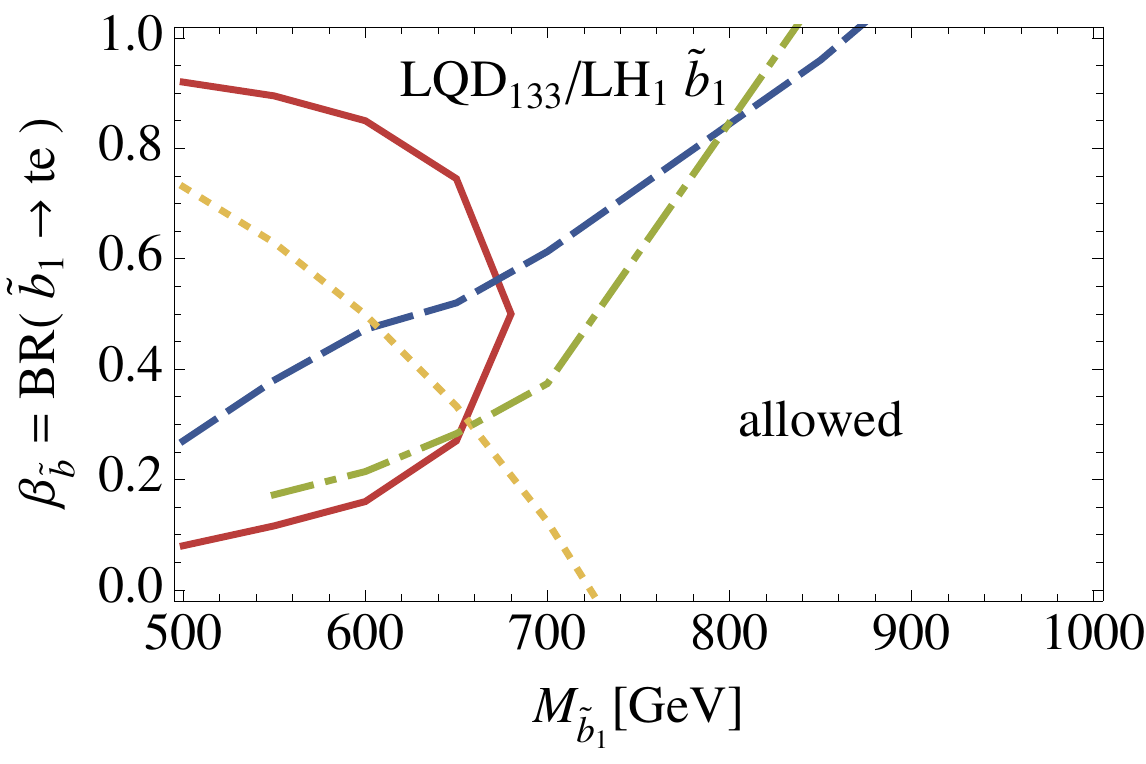}
\includegraphics[width=0.49\textwidth]{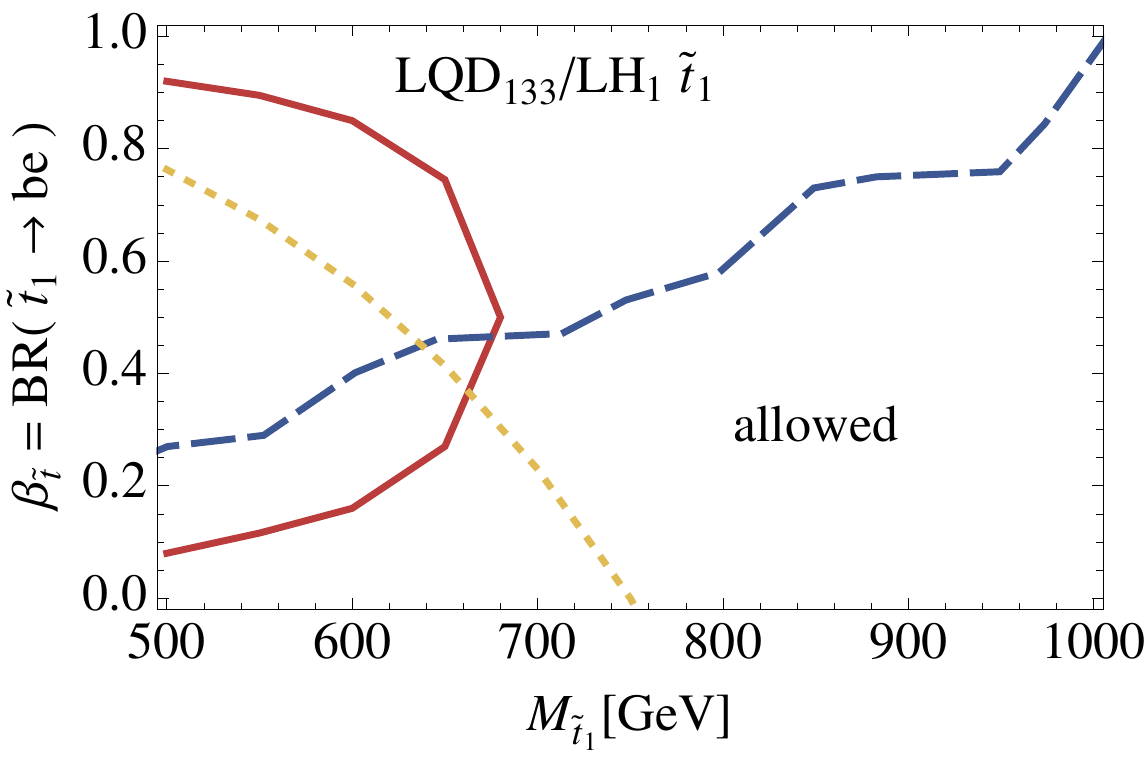}
\caption{95\%C.L. Exclusion plots for the sbottom LSP ({\bf left}) and the stop LSP ({\bf right}) from CMS leptoquark searches in $eejj$ (blue-dashed) and $e\nu jj$ (red-solid) channels. Also shown are CMS RPC sbottom and stop searches (yellow-dotted) in $b\bar{b}$+MET and $t\bar{t}$+MET channels. For sbottoms, CMS multilepton ($\geq 3\ell$) RPV search constrains additionally (green-dot-dashed). The region left to each line is excluded. The bounds are equally applicable to LH$_1$ and LQD$_{133}$ models.}
\label{fig:bound1}
\end{figure}

Let us first consider how the sbottom LSP can be constrained at the LHC. 
Sbottom pair productions in the  LH$_1$ and LQD$_{133}$ models, leave the final states:  
\beq 
\widetilde{b}_1 \widetilde{b}_1^* \, \to  \, bb \nu \nu, \, tbe\nu,  \, tt ee.
\eeq
The $bb\nu\nu$ is constrained by RPC sbottom searches through $\widetilde{b}_1 \to b \chi_1^0$ with the massless LSP, hence $b\bar{b}+$missing transverse energy(MET). The existing strongest bound on the sbottom mass is 725GeV from CMS 19.4/fb~\cite{CMS:2014nia}. The $tbe\nu$ can be constrained from the $e\nu jj$ searches of first-generation leptoquarks~\cite{cms-LQ} -- the CMS analysis uses two hardest jets of any flavor. Note that the sbottom and the leptoquark have the same quantum numbers as color triplet, and their production rates are almost identical, as dictated by QCD interactions. So it is appropriate to use this result to extract bounds on sbottoms. The $ttee$ can be constrained from the $eejj$ searches of leptoquarks and additionally from multi-lepton($\geq 3\ell$) RPV LLE searches~\cite{Chatrchyan:2013xsw}. We comment on other searches in Appendix~\ref{app:bound}.

We recast these search results to exclusion bounds on the sbottom in the left panel of \Fig{fig:bound1} -- we refer to Appendix~\ref{app:bound} for how we obtain these bounds. The same bounds apply to both LQD$_{133}$ and LH$_1$ as they predict the same final states. Large $\beta_{\tilde b}$ is constrained from the $eejj$ and the multi-lepton RPV searches whereas small $\beta_{\tilde b}$ is constrained from the RPC sbottom search. In general, sbottoms lighter than about 660 GeV is excluded by at least one of those searches.

\medskip \medskip

\begin{figure}[t] 
\includegraphics[width=0.49\textwidth]{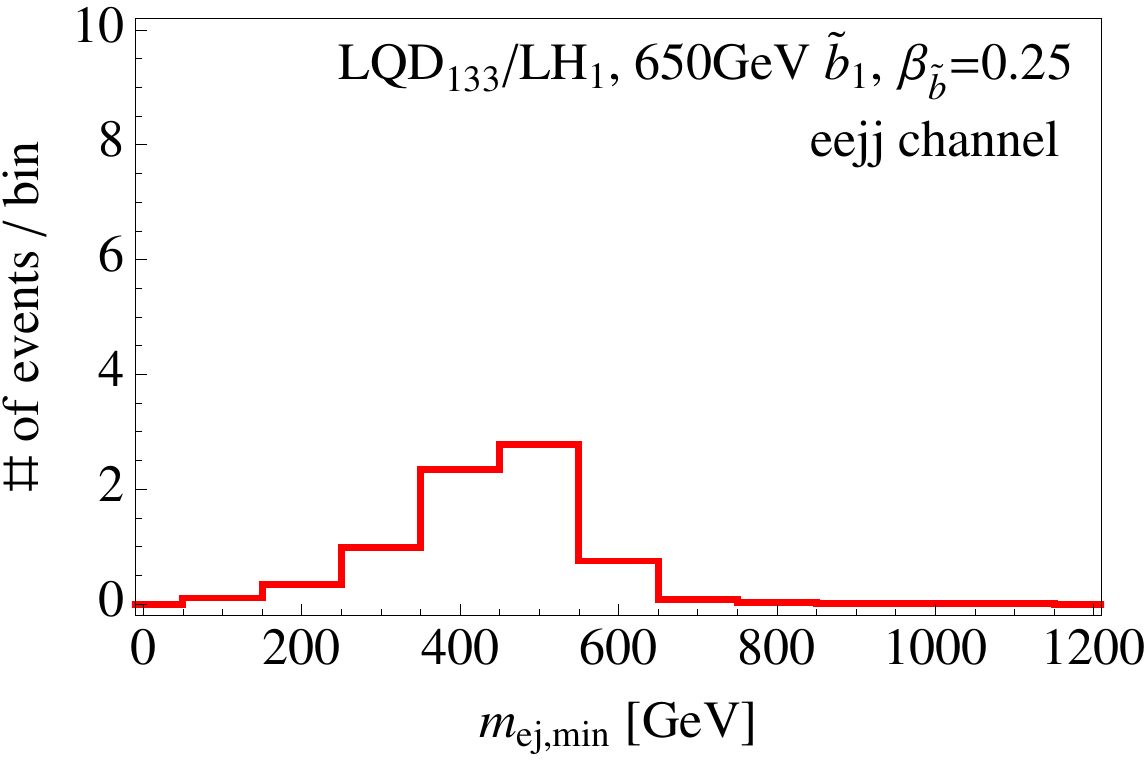}
\includegraphics[width=0.49\textwidth]{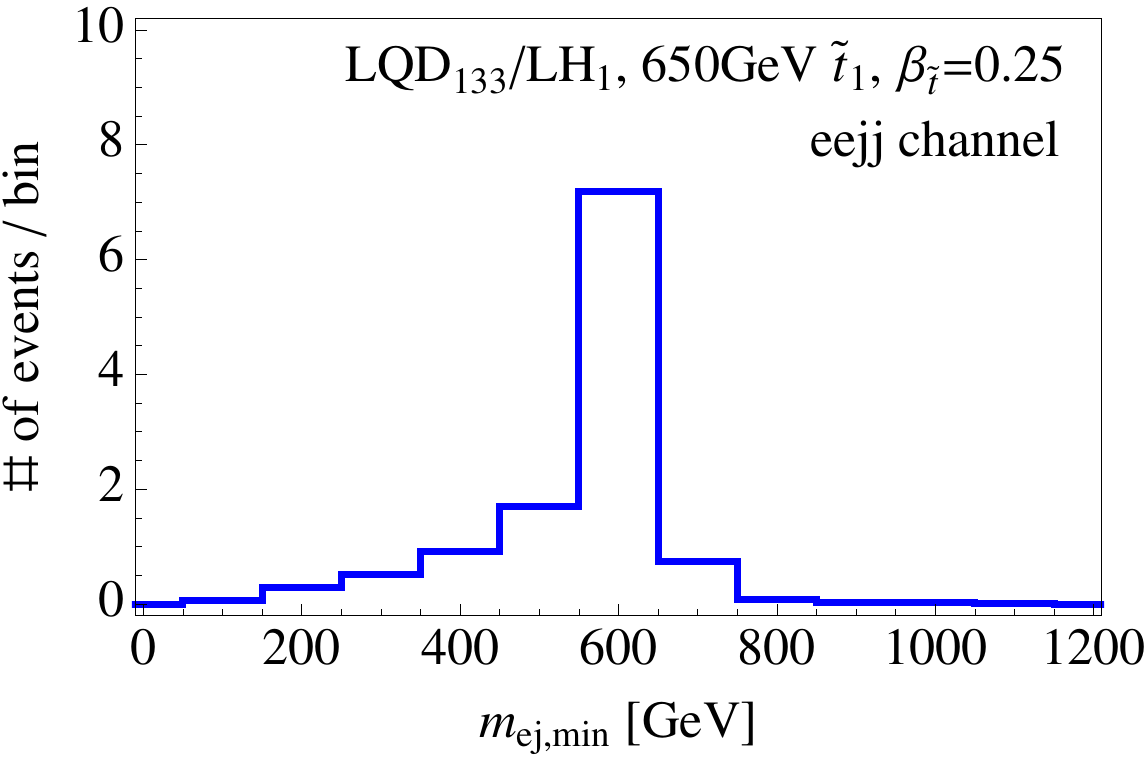}
\caption{Invariant mass, $m_{ej,{\rm min}}$, from 650GeV sbottom({\bf left}) or stop({\bf right}) pairs decaying to $ee jj$ channel via LH$_1$ or LQD$_{133}$ RPV couplings. CMS leptoquark search cuts are applied except for the cut on the invariant mass. 19.6/fb is assumed. $\beta=0.25$ is chosen for illustration.} 
\label{fig:mej}
\end{figure}

We now turn to the stop LSP. Stop pairs in the LH$_1$ and LQD$_{133}$ models decay as 
\beq
\widetilde{t}_1 \widetilde{t}_1^*  \, \to  \, tt \nu \nu, \, tb e\nu, \, bb ee,
\eeq
where the first two modes are not allowed in the  LQD$_{133}$ model.
The $tt\nu\nu$ channel is constrained by RPC stop searches through $\widetilde{t}_1 \to t \chi_1^0$ with the massless LSP. The existing strongest bound is 750\,GeV from CMS 19.5/fb~\cite{CMS:2014wsa}. The remaining decay modes, $tbe\nu$ and $bbee$, can be constrained from the $e\nu jj$ and $eejj$ searches of first-generation leptoquarks~\cite{cms-LQ}. Note that the stop also has the same quantum numbers as leptoquarks. Unlike sbottoms, stop pairs 
do not lead to final states with more than 3 leptons.
Recasting these search results to exclusion bounds on the stop, we obtain the right panel of \Fig{fig:bound1}. 
Similarly to the sbottom case, stops lighter than about 660 GeV is excluded.

\medskip

There is one notable difference between the sbottom LSP and the stop LSP. Sbottom pairs decay to $ttee$ while stop pairs decay 
to $bbee$. Tops  produce  more jets, and each jet becomes softer as decay products share the energy-momentum of sbottoms. Thus the acceptance under leptoquark search cuts gets lower. The $eejj$ exclusion bound (blue-dashed) on sbottoms 
(the left panel of \Fig{fig:bound1}) is indeed weaker than that 
on stops (the right panel of \Fig{fig:bound1}). Likewise, the $e\nu jj$ bound (red solid) in \Fig{fig:bound1} is also weaker than 
the official $e\nu jj$ bound on the leptoquark model in Ref.~\cite{cms-LQ}.

Most notably, the invariant mass of the $ej$ pair, $m_{ej,{\rm min}}$, does not reconstruct the sbottom mass. In \Fig{fig:mej}, we contrast the invariant mass spectrum for the sbottom LSP and the stop LSP. We choose the presumably correct $ej$ pair according to the CMS leptoquark analysis; the pair giving smaller invariant mass difference is selected. The $m_{ej}$ from sbottoms have a broader spectrum and the peak formed at a lower mass because not all top decay products are included. 
It will be useful to measure this characteristic difference in the future searches.

Therefore, potentially significant improvements in the third-generation squark LSP searches can be achieved with $b$-taggings and/or top reconstructions. With 20/fb of data, 8.1fb $\times$ 20/fb $\simeq$ 160 pairs of 700 GeV sbottoms are produced, and much better bounds are beginning to be statistically limited. In any case, 160 is still a reasonably large number, and more dedicated searches implementing $b$-tagging and/or top reconstruction are certainly worthwhile.

\medskip 

\begin{figure}[t] 
\includegraphics[width=0.49\textwidth]{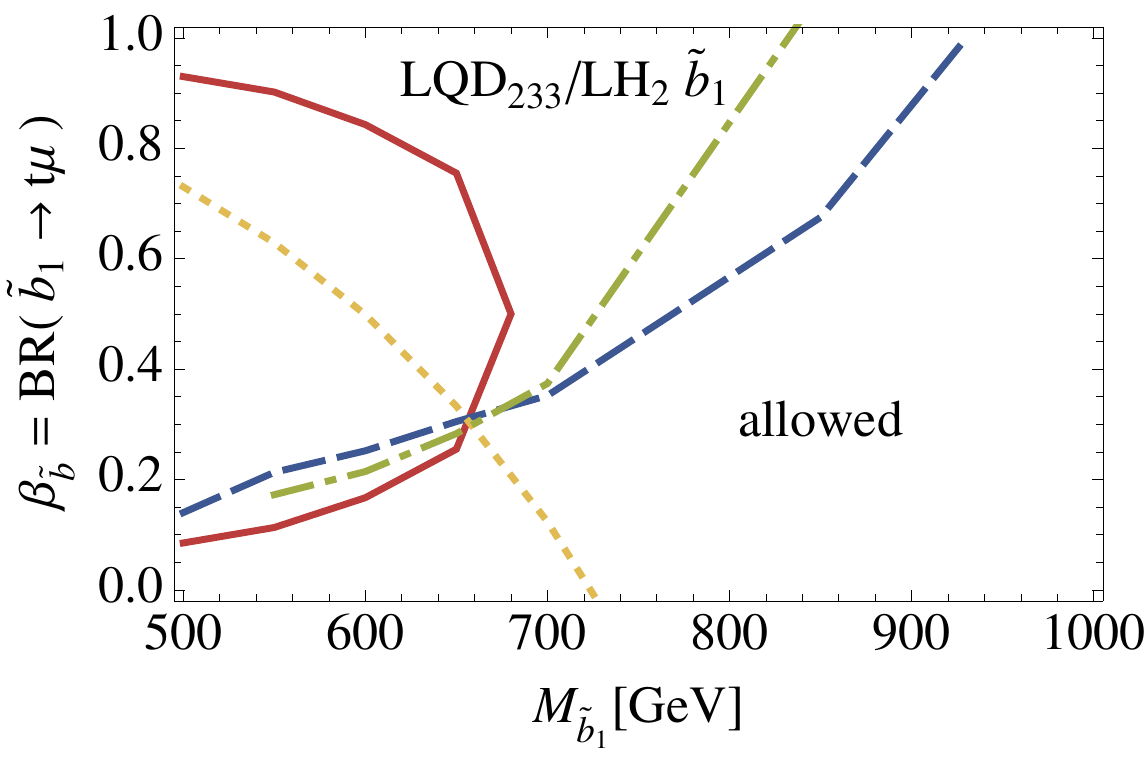}
\includegraphics[width=0.49\textwidth]{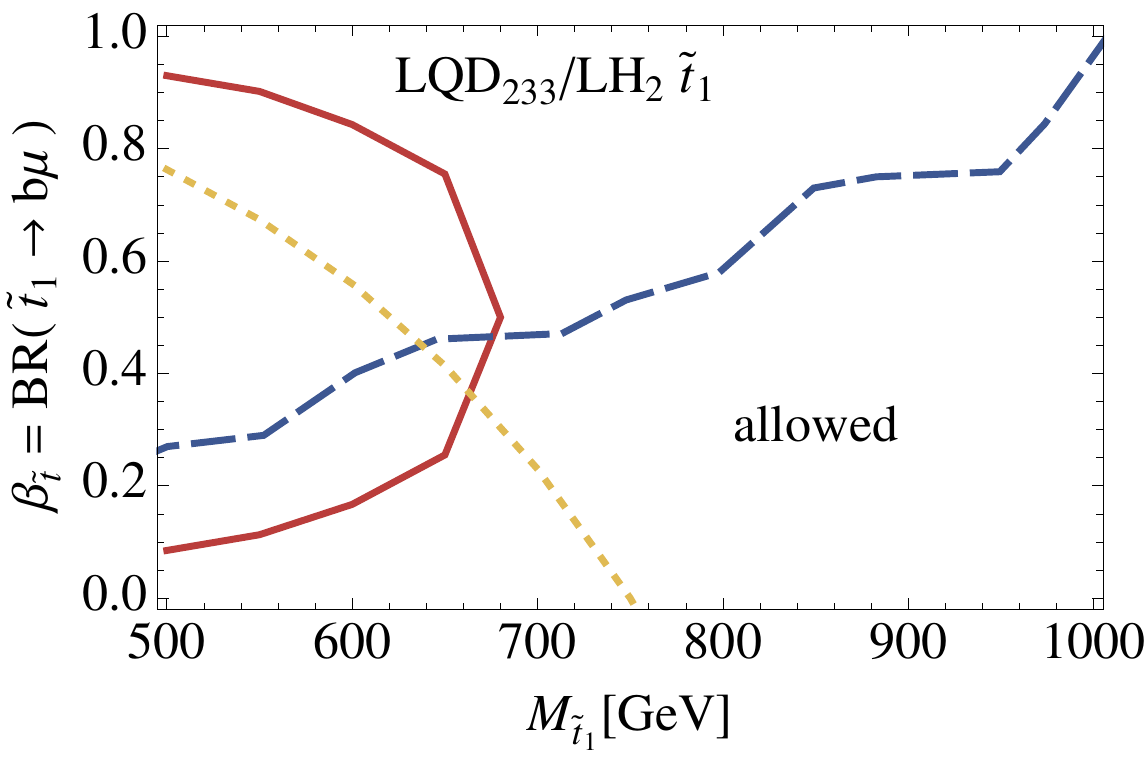}
\caption{Same as in \Fig{fig:bound1} but the $\mu$ channel results of CMS leptoquark searches~\cite{cms-LQ2} are used 
for the red-solid and blue-dashed lines, which constrain the models LH$_2$ and LQD$_{233}$ models.}
\label{fig:bound2}
\end{figure}

We can repeat the same analysis in the  LH$_2$ and LQD$_{233}$ models allowing sbottom and stop LSP decays to $\mu$, and 
apply  the CMS second-generation leptoquark searches~\cite{cms-LQ2}.  The resulting bounds are shown  in \Fig{fig:bound2}.
 Compared to the $eejj$ search, the $\mu\mu jj$ search is somewhat more stringent partly because $\mu$ is more accurately measured and cleaner -- compare blue-dashed lines in the left panels of \Fig{fig:bound1} and \Fig{fig:bound2}. On the other hand, $\mu \nu jj$ results are similar to $e \nu jj$ results  (red-solid lines).  To summarize, again,
third-generation squark LSPs lighter than about 660 GeV are generally excluded.

\medskip

Finally, the LQD$_{ij3+i3j}$ models with $i, j=1,2$ are equivalent to the leptoquark  models and the current search results can be directly applied to constrain the sbottom/stop LSP mass.

\section{The Observed Leptoquark Excess from Sbottom Decays}
\label{sec:LQ}

The CMS leptoquark analysis has recently reported excesses in 650GeV leptoquark searches in both $eejj$ and $e\nu jj$ channels~\cite{cms-LQ}. The excesses are claimed to be 2.4 and 2.6$\sigma$ significant, respectively. The excesses disappear when a $b$-jet is required, and no similar excess is observed in searches with $\mu$~\cite{cms-LQ2} and $\tau$~\cite{cms-LQ3}. In this section, we discuss how our third model, LQD$_{113+131}$, can fit the excesses.

\begin{figure}[t] 
\includegraphics[width=0.49\textwidth]{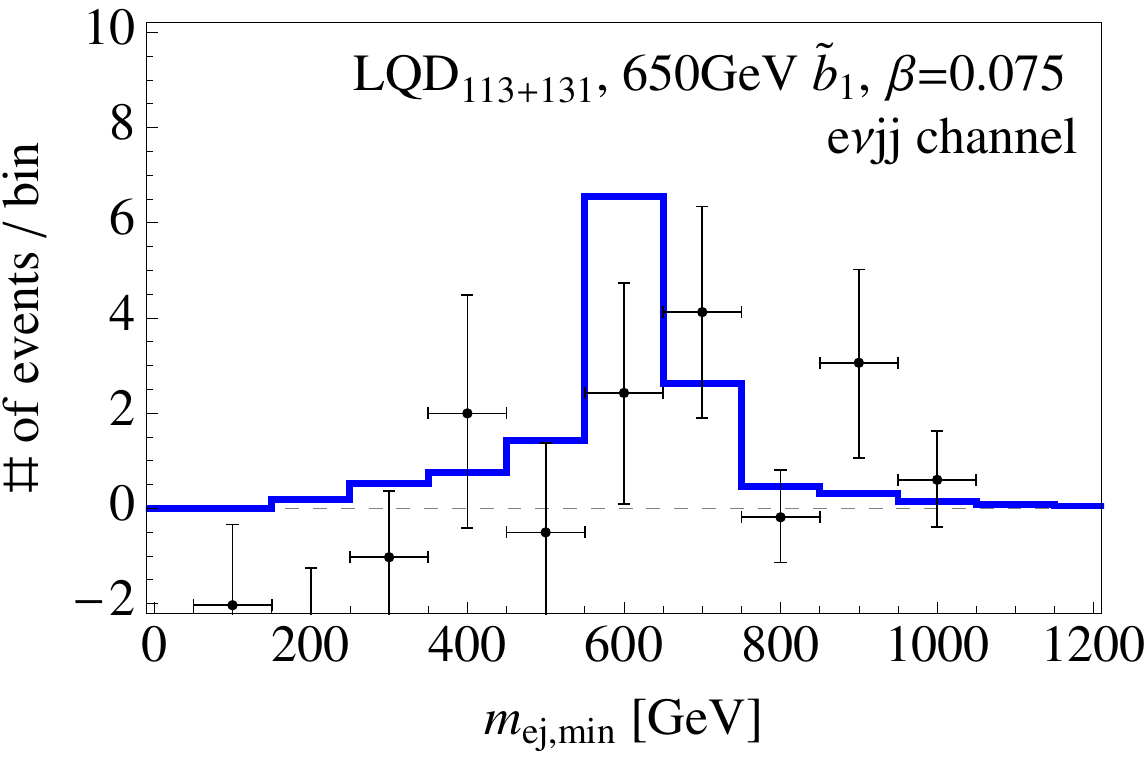}
\caption{The invariant mass, $m_{ej,{\rm min}}$, from 650GeV sbottom pairs decaying to the $e\nu jj$ channel 
via LQD$_{113+131}$ RPV couplings. CMS leptoquark search cuts are applied except for the cut on the invariant mass. 19.6/fb is assumed. A small $\beta=0.075$ giving a good fit to data is chosen. The data with SM predictions subtracted are taken from CMS results in Ref.~\cite{cms-LQ}.} 
\label{fig:mej2}
\end{figure}

\subsection{Sbottoms as Leptoquarks}
Sbottom pairs in the LQD$_{113+131}$ model decay as
\beq
\widetilde{b}_1 \widetilde{b}_1^* \, \to \, dd \nu \nu, \, du e \nu , \, uu ee,
\eeq
with BR=$(1-\beta)^2, \, 2 \beta(1-\beta)$ and $\beta^2$, respectively. This model is identical to the first-generation leptoquark model considered in the CMS analysis except that $\beta$ is given differently by \Eq{eq:beta-leptoquark} in our model. The best fit is allegedly reported to be with 650\,GeV and $\beta=0.075$. Our model can accommodate this by the decay of sbottom LSPs. By simply assuming $\lambda^\prime_{113} = \lambda^\prime_{131}$ as an example, we can extract more specific information on the underlying parameters. Then,  $\beta = \sin^2 \theta_{\tilde b} / (1+\sin^2 \theta_{\tilde b} ) \leq 0.5$ is now bounded from above. The best-fit value, $\beta=0.075$, requires $\sin^2 \theta_{\tilde b} = 0.081$, meaning that the sbottom LSP is mostly left-handed. The constraint from electroweak precision test is briefly discussed in the next subsection. 

The $m_{ej,{\rm min}}$ invariant mass spectrum is also scrutinized in the CMS analysis. So far, no sharp peak is observed unlike the expectation from leptoquark decays. As compared to our previous two models, the LQD$_{113+131}$ does not involve top quarks and would also predict the same sharp peak in the invariant mass as leptoquark model does. See \Fig{fig:mej2} for the comparison of the model prediction and data -- no clear resonance-like structure is seen in data, but the model prediction is not significantly different from data yet.

\medskip

Our interpretation of the sbottom LSP in the LQD$_{113+131}$ model as a leptoquark of 650 GeV responsible for the mild CMS excesses  requires the corresponding couplings, $\lambda'_{113}$ and $\lambda'_{131}$, to dominate over other sbottom LSP RPV couplings if any. As discussed in Eq.~(\ref{xi-mnu},\ref{lambda-mnu}), these couplings can take the values of  $\lambda'_{113} \sim  \lambda'_{131} \sim  10^{-3}$ to produce (mainly) the (11) component of the observed neutrino mass matrix.  Then, the other components can come from smaller bilinear RPV couplings $\xi_i c_\beta \sim 10^{-6}$ and/or trilinear couplings, e.g.,
$\lambda'_{i33} \sim 10^{-4}$ to produce $m^{\rm tree}_{\nu, ij} \propto \xi_i \xi_j c_\beta^2$ and/or $ m^{\rm loop}_{\nu, ij} \propto \lambda'_{i33} \lambda'_{j33}$. 
In this scenario, the sbottom LSP can have additional but suppressed decay modes in the $\mu$ and $\tau$ channels which may provide a test of the model.  Of course, the neutrino mass components can come mainly from the LLE couplings,
e.g.,  $m^{\rm loop}_{\nu, ij} \propto  \lambda_{i33} \lambda_{j33}$, which has no impact on
 the sbottom LSP phenomenology.

\subsection{Electroweak Precision Data and Stop Masses}

\begin{figure}[t] 
\includegraphics[width=0.49\textwidth]{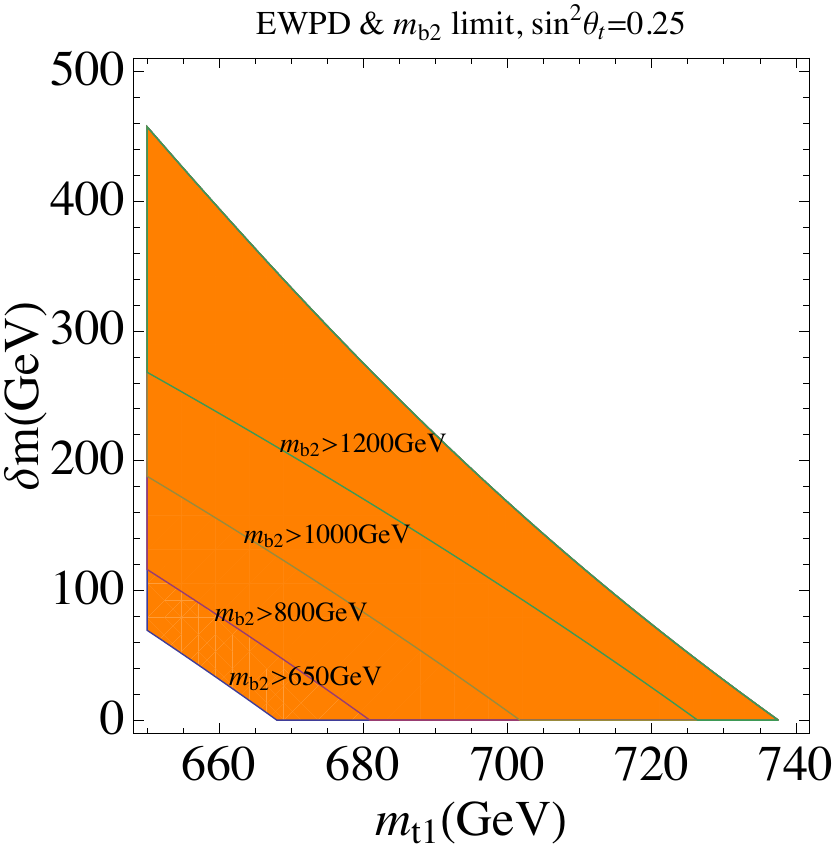}
\includegraphics[width=0.49\textwidth]{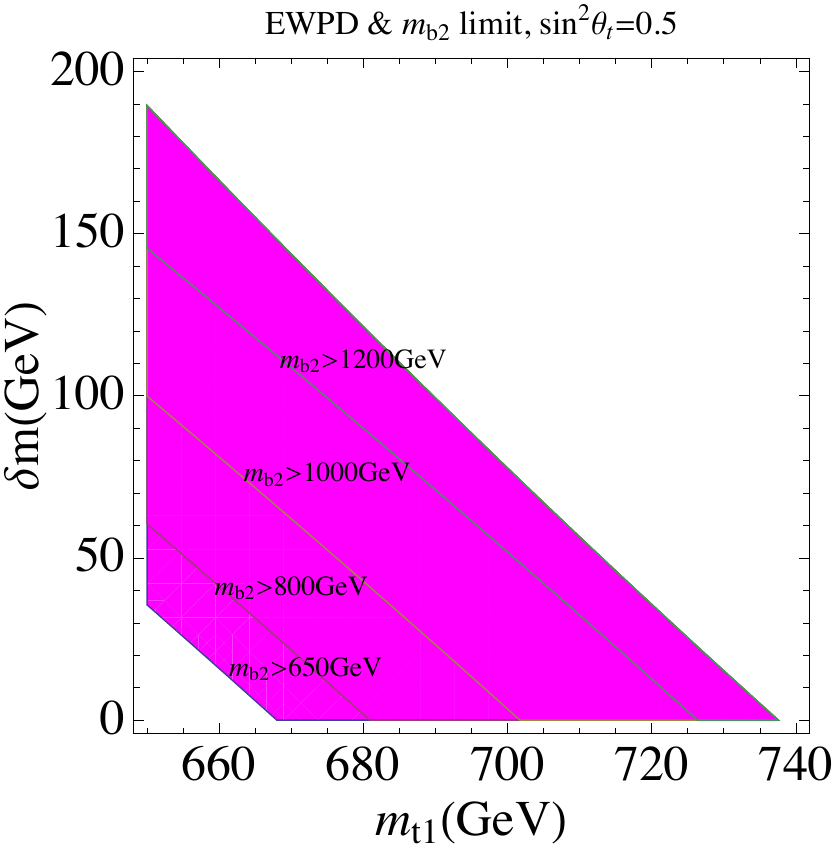}
\caption{The EWPD constraints on stop and heavy sbottom masses for the best-fit parameters, with 650\,GeV sbottom LSP and $\sin^2 \theta_{\tilde b}=0.081$. Here, $\delta m  \equiv m_{{\tilde t}_2} - m_{{\tilde t}_1} $ is the stop mass splitting.
Lighter stop mass is bounded by EWPD and the stop mass splitting scales up as the bound on the heavier sbottom mass increases. In both figures, $\tan\beta=10$ is chosen. } 
\label{fig:ewpd}
\end{figure}

The mostly left-handed sbottom solution obtained in the previous subsection may imply that other stops (and/or sbottoms) are also light; otherwise, the model is inconsistent with the electroweak precision data(EWPD). The possible other light particles can provide additional collider constraints on the model. 
Indeed, it has been shown that the EWPD can give important constraints on the stop masses and mixing angles in combination with the RPC searches of sbottoms~\cite{sbottom}. 

The deviation from the custodial symmetry in the SM is bounded to \cite{susyewpt}
\bea
(\Delta \rho_0)^{\pm}&=&(\rho_0)_{m_h=125\,{\rm GeV}} -1 \nonumber \\
&=&(4.2\pm 2.7) \times 10^{-4}.
\eea 

The sbottom and stop contribution to the $\rho$ parameter \cite{susyrho} is
\bea
\Delta \rho^{SUSY}_0&=& \frac{3G_\mu}{8\sqrt{2} \pi^2} \bigg[ -\sin^2\theta_{\tilde t} \cos^2\theta_{\tilde t} F_0 (m^2_{{\tilde t}_1},m^2_{{\tilde t}_2}) -\sin^2\theta_{\tilde b} \cos^2\theta_{\tilde b} F_0 (m^2_{{\tilde b}_1},m^2_{{\tilde b}_2}) \nonumber \\
&&\quad \quad +\cos^2\theta_{\tilde t} \cos^2\theta_{\tilde b} F_0 (m^2_{{\tilde t}_1},m^2_{{\tilde b}_1}) +\cos^2\theta_{\tilde t} \sin^2\theta_{\tilde b} F_0 (m^2_{{\tilde t}_1},m^2_{{\tilde b}_2}) \nonumber \\
&&\quad \quad + \sin^2\theta_{\tilde t} \cos^2\theta_{\tilde b} F_0 (m^2_{{\tilde t}_2},m^2_{{\tilde b}_1}) + \sin^2\theta_{\tilde t} \sin^2\theta_{\tilde b} F_0 (m^2_{{\tilde t}_2},m^2_{{\tilde b}_2}) \bigg] ,
\eea
where $F_0$ is defined by
\beq
F_0[x,y] = x+y-\frac{2xy}{x-y} \,\log \frac{x}{y}.
\eeq
From the mass terms for stops and sbottoms, we can infer the following relation between physical squark masses and mixing angles,
\bea
\sin^2\theta_{\tilde b}\,m^2_{{\tilde b}_2}+\cos^2\theta_{\tilde b} \,m^2_{{\tilde b}_1}&=& \cos^2\theta_{\tilde t} \,m^2_{{\tilde t}_1} +\sin^2\theta_{\tilde t} \,m^2_{{\tilde t}_2}-m^2_t  -m^2_W \cos(2\beta) + m^2_b .
\eea

In \Fig{fig:ewpd}, we show bounds on the masses of other sbottoms and stops by assuming the best-fit parameters, $m_{\tilde b_1} = 650$ GeV and $\sin^2 \theta_{\tilde b}=0.081$, chosen in the previous subsection.  Although the EWPD bound depends on various other parameters including stop mixing angle, the lighter stop mass is bounded up to about $740\,{\rm GeV}$  and the stop mass splitting is bounded up to about $190\,{\rm GeV}$ for a maximal stop mixing. 
In particular, when the collider limit on the heavier sbottom mass increases, the lighter stop mass and the stop mass splitting tend to get larger so the allowed parameter space in the stop sector is reduced.

The $125\,{\rm GeV}$ Higgs mass would require stop masses of $500-800\,{\rm GeV}$ for a maximal stop mixing or stop masses above $3\,{\rm TeV}$ for a zero stop mixing \cite{hall}. 
Thus, in the case of a small stop mixing, the Higgs mass condition would be incompatible with EWPD. 
On the other hand, for a maximal stop mixing, the stop masses required for the Higgs mass can constrain the parameter space further. When there is a new dynamics for enhancing the Higgs mass such as a singlet chiral superfield, 
we may take the EWPD in combination with sbottom mass limit to be a robust bound on stop masses.

\section{Summary and Conclusion}
\label{sec:summary}

Through LH and LQD RPV couplings, the third-generation squark LSP can decay to leptons and jets. Jet+MET final states are constrained by conventional RPC SUSY searches, and multilepton(+jets)+MET final states are constrained from leptoquark searches as well as multilepton RPV searches.  We found that the sbottom and the stop LSP decaying to $e$ or $\mu$ are similarly well constrained up to about 0.66 $\sim$ 1 TeV depending on leptonic branching fractions. When the sbottom decays to a top quark and an electron as in the LH$_i$ and LQD$_{i33}$ models, the bounds are slightly weaker as each top decay product is softer and not all is used in the analyses. The resulting characteristically different $m_{ej}$ invariant mass spectra can distinguish the models. More dedicated search for this case can be pursued by implementing $b$-taggings and/or top reconstructions. The bounds on $\mu$ final states are somewhat stronger than those on $e$ final states so that a wider region of parameter space above 660 GeV is excluded for the LH$_2$ and LQD$_{233}$ models. Lastly, we proposed the LQD$_{113+131}$ model with sbottom LSPs as a good fit to the recently 
observed mild leptoquark excesses and discussed its possible implications on the masses of other stops and sbottoms in view of the EWPD and the 125 GeV Higgs mass.

\vspace{6mm}
{\it Acknowledgement.}
We thank Suyong Choi  for helpful comments.
E.J.C is supported by the NRF grant funded by the Korea government (MSIP) (No.\ 2009-0083526) through 
KNRC at Seoul National University.
S.J. is supported in part by the NRF grant (2013R1A1A2058449).
H.M.L is supported in part by Basic Science Research Program through the NRF grant (2013R1A1A2007919) and by the Chung-Ang University Research Grants in 2014. S.C.P is supported by Basic Science Research Program through the NRF grants (NRF-2011-0029758 and NRF-2013R1A1A2064120).

\appendix

\section{Approximate Diagonalizations of RPV Masses} \label{app:diag}

Bilinear RPV in superpotential and soft SUSY breaking scalar potential leads to the mixing between neutrinos (charged leptons)
and neutralinos (charginos).  As such bilinear couplings are required to be small to produce tiny neutrino masses, it is 
convenient to rotate  away first these mixing masses by the following approximate diagonalizations collected from Ref.~\cite{chun02}.

(i) Neutrino--neutralino diagonalization:
\begin{equation}
 \begin{pmatrix} \nu_i \cr \chi^0_j \end{pmatrix}  \longrightarrow
 \begin{pmatrix} \nu_i- \theta^N_{ik} \chi^0_k  \cr
           \chi^0_j + \theta^N_{lj} \nu_l \end{pmatrix},
\end{equation}
where $(\nu_i)$ and $(\chi^0_j)$ represent three neutrinos
$(\nu_e, \nu_\mu, \nu_\tau)$ and four neutralinos
$(\tilde{B}, \tilde{W}_3, \tilde{H}^0_d, \tilde{H}^0_u)$ 
in the flavor basis, respectively.  The rotation elements
$\theta^N_{ij}$ are given by 
\begin{eqnarray}
 \theta^N_{ij} &=& \xi_i c^N_j c_\beta - \epsilon_i \delta_{j3} 
 \quad\mbox{and} \\
 (c^N_j) &=& {M_Z \over F_N} ({ s_W M_2 \over c_W^2 M_1 + s_W^2 M_2},
  -{ c_W M_1 \over c_W^2 M_1 + s_W^2 M_2}, -s_\beta{M_Z\over \mu},
   c_\beta{M_Z\over \mu}), \nonumber
\end{eqnarray}
where $\xi_i \equiv a_i - \epsilon_i$ and   $F_N=M_1 M_2 /( c_W^2 M_1 + s_W^2 M_2) + M_Z^2 s_{2\beta}/\mu$.
Here  $s_W=\sin\theta_W$ and $c_W=\cos\theta_W$ 
with the weak mixing angle $\theta_W$.  

(ii) Charged-lepton--chargino diagonalization:
\begin{equation}
 \begin{pmatrix} e_i \cr \chi^-_j \end{pmatrix} \rightarrow
 \begin{pmatrix}  e_i- \theta^L_{ik} \chi^-_k  \cr
           \chi^-_j + \theta^L_{lj} e_l \end{pmatrix} \quad;\quad
 \begin{pmatrix} e^c_i \cr \chi^+_j \end{pmatrix} \rightarrow
 \begin{pmatrix} e^c_i- \theta^R_{ik} \chi^+_k  \cr
           \chi^+_j + \theta^R_{lj} e^c_l \end{pmatrix}, 
\end{equation}
where $e_i$ and $e^c_i$ denote the left-handed charged leptons and
anti-leptons, $(\chi^-_j)=(\tilde{W}^-,\tilde{H}^-)$ and 
$(\chi^+_j)=(\tilde{W}^+,\tilde{H}^+)$.
The rotation elements $\theta^{L,R}_{ij}$ are  given by
\begin{eqnarray}
&& \theta^L_{ij}= \xi_i c^L_j c_\beta-\epsilon_i \delta_{j2}\;, \quad
 \theta^R_{ij}= {m^e_i\over F_C} \xi_i c^R_j c_\beta  \quad\mbox{and} \\
&&  (c^L_j)= -{M_W \over F_C} (\sqrt{2}, 2s_\beta{M_W\over \mu})\;,
               \nonumber \\
&&   (c^R_j)= -{M_W  \over F_C} (\sqrt{2}(1-{M_2\over \mu} t_\beta), 
      \frac{M_2^2 c^{-1}_\beta}{\mu M_W }+2{M_W \over \mu} c_\beta), 
      \nonumber
\end{eqnarray}
and $F_C= M_2 + M_W^2 s_{2\beta}/\mu$.

\section{Bound Estimation} \label{app:bound}

Here we summarize how we reinterpret LHC results to obtain exclusion bounds on our models. We use the next-to-leading order sbottom production cross sections in Refs.~\cite{sbottomxection,susyprod}. 

The $bb\nu \nu$ final states are constrained from RPC sbottom pair searches. Sbottoms decaying to $b \chi_1^0$ 100\% is currently limited to be above 720GeV~\cite{CMS:2014nia}. For our given sbottom mass, ignoring differences in cut efficiencies and kinematics, we find the branching ratio suppression needed to make the production rate of the given $\widetilde{b}_1 \widetilde{b}_1^* \to b b \nu \nu$ equal to that of 720GeV sbottom pairs. We reinterpret the stop RPC searches in the $t\bar{t}$+MET channel~\cite{CMS:2014wsa} in the same way to constrain $tt \nu \nu$ final states. For the LQD$_{113+131}$, the RPC searches of squark pairs can be similarly relevant. Interestingly, a \emph{single} squark pair is weakly constrained from the $q\bar{q}$+MET search~\cite{CMS:2014ksa} to be above only 570 GeV -- but they can still exclude small part of surviving parameter space. 

Various $\ell \nu jj$ final states are constrained from leptoquark searches. Leptoquark searches~\cite{cms-LQ, cms-LQ2} display several set of cuts(signal regions) optimized for different leptoquark masses. For the LQD$_{113+131}$ which have exactly the same kind of decay modes as leptoquarks, the official CMS exclusion bounds on leptoquarks apply equally well. For the LQD$_{i33}$ and LH$_i$ models which involve heavy quarks in the final states, we carry out Monte-Carlo simulations (based on \texttt{MadGraph}~\cite{Alwall:2011uj}, \texttt{Pythia}~\cite{Sjostrand:2006za} and \texttt{FastJet}~\cite{Cacciari:2011ma}), estimate efficiencies under all displayed cuts and use the most constraining result. To quantify the deviation, we add statistical error, $\sqrt{S+B}$, and the reported systematic errors in quadrature -- our own 95\%C.L.$\simeq 1.96\sigma$ exclusion bounds on leptoquarks based on this method agree well with the official results. As different signal regions are not mutually exclusive, we do not $\chi^2$ them.

The $ttee$ final states can involve more than three leptons or same-sign dileptons and $b$-jets which are often clean. We find that multilepton($N_\ell \geq 3$) RPV LLE search~\cite{Chatrchyan:2013xsw} with various binned discovery cuts is most relevant to us. We simulate all the discovery cuts with $300 < S_T < 1500$ GeV and use the most stringent result to obtain bounds. The strongest bound is usually from discovery cuts with $\geq 1b$ and $S_T \gtrsim 1000$ GeV requirements. Similar searches of same-sign dileptons plus $b$-jets plus multijets~\cite{Aad:2014pda}, four-lepton~\cite{Aad:2014iza} and other $\geq 3\ell$ + $b$-jet searches in, e.g., Refs.~\cite{TheATLAScollaboration:2013jha} are less optimized for our benchmark models of about 700GeV squarks.


\end{document}